\documentclass[a4paper, 10pt, conference]{ieeeconf}      

\IEEEoverridecommandlockouts                              

\usepackage{cite}
\usepackage{amsmath,amssymb,amsfonts}
\usepackage[left=15.9mm,right=15.9mm,top=19.1mm,bottom=25.4mm]{geometry}
\usepackage{algorithmic}
\usepackage{graphicx}
\usepackage{booktabs}
\usepackage{textcomp}
\usepackage{xcolor}
\usepackage{array}
\usepackage{multirow}
\usepackage{float}
\usepackage{caption}
\usepackage{subcaption}
\usepackage{footnote}
\usepackage{hyperref} 
\hypersetup{
    linkbordercolor={0 1 0}
}

\title{\LARGE\bf
Using Explainable AI for EEG-based Reduced Montage Neonatal Seizure Detection
}

\author{Dinuka Sandun Udayantha$^{1}$\authorrefmark{1}, Kavindu Weerasinghe$^1$, Nima Wickramasinghe$^1$, Akila Abeyratne$^1$,\\ Kithmin Wickremasinghe$^2$, Jithangi Wanigasinghe$^3$, Anjula De Silva$^1$, and Chamira U. S. Edussooriya$^1$\authorrefmark{2}\\[5pt]
$^1$Department Electronic and Telecommunication Eng., University of Moratuwa, Sri Lanka\\
$^2$Department Electrical and Computer Eng., University of British Columbia, Canada\\
$^3$Department of Paediatrics, Faculty of Medicine, University of Colombo, Sri Lanka\\[5pt]
\textit{email: \authorrefmark{1}udayanthabds.19@uom.lk , \authorrefmark{2}chamira@uom.lk}
\thanks{\copyright 20XX IEEE.  Personal use of this material is permitted.  Permission from IEEE must be obtained for all other uses, in any current or future media, including reprinting/republishing this material for advertising or promotional purposes, creating new collective works, for resale or redistribution to servers or lists, or reuse of any copyrighted component of this work in other works.}
}

\begin{document}

\maketitle
\thispagestyle{empty}
\pagestyle{empty}

\begin{abstract}

The neonatal period is the most vulnerable time for the development of seizures. Seizures in the immature brain lead to detrimental consequences, therefore require early diagnosis. The gold-standard for neonatal seizure detection currently relies on continuous video-EEG monitoring; which involves recording multi-channel electroencephalogram (EEG) alongside real-time video monitoring within a neonatal intensive care unit (NICU). However, video-EEG monitoring technology requires clinical expertise and is often limited to technologically advanced and resourceful settings. Cost-effective new techniques could help the medical fraternity make an accurate diagnosis and advocate treatment without delay. In this work, a novel explainable deep learning model to automate the neonatal seizure detection process with a reduced EEG montage is proposed, which employs convolutional nets, graph attention layers, and fully connected layers. Beyond its ability to detect seizures in real time with a reduced montage, this model offers the unique advantage of real-time interpretability. By evaluating the performance on the Zenodo dataset with 10-fold cross-validation, the presented model achieves an absolute improvement of 8.31\% and 42.86\% in area under curve (AUC) and recall, respectively.

\end{abstract}

\section{Introduction}
\label{sec:intro}

\begin{figure*}[!ht]
    \centering
    \includegraphics[scale=0.31]{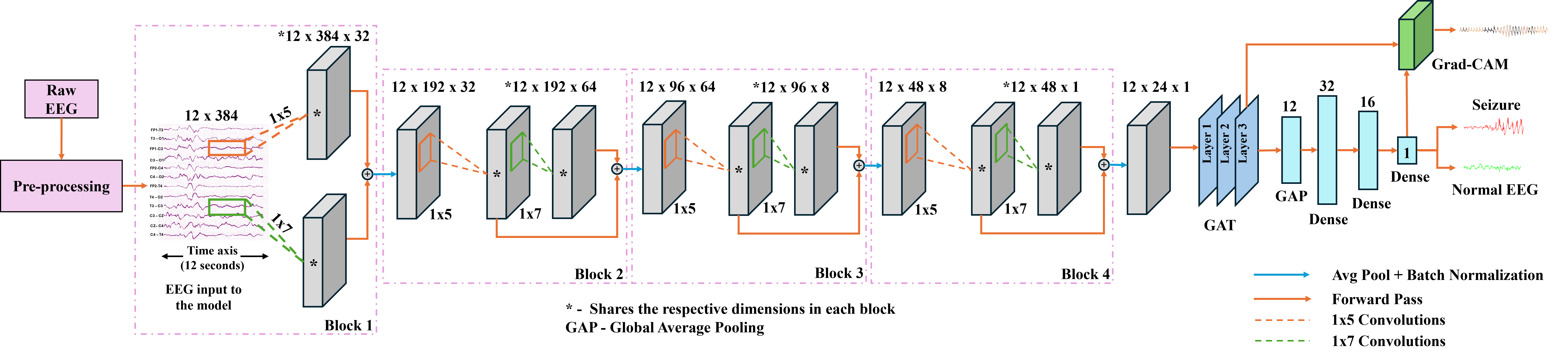}
    \caption{The proposed deep learning model architecture. The first $4$ blocks belong to the CNN-based temporal feature extractor. The $12$ channels are preserved throughout the CNN encoder but down-sampled the temporal features at the end of each block using Average Pooling. GAT layers $1, 2,$ and $3$ have the output shapes $(12\times 37)$, $(12\times 32)$, and $(12\times 16)$ respectively and the last multilayer perceptron (MLP) network has $32, 16$, and $1$ neurons, respectively.}
    \label{fig:CNN}
    \vspace{-15pt}
\end{figure*}

Neonatal seizures are epileptic seizures that occur in infants that are younger than $4$ weeks. In an EEG, it can be identified as an occurrence of sudden, abnormal, and paroxysmal ictal rhythm with a $2$ $\mu V$ or higher amplitude. This 4-week neonatal period is the most vulnerable time to develop seizures, capable of causing significant harm to the developing brain and necessitating prompt diagnosis followed by treatment. According to Kang \textit{et al.} \cite{kang2015neonatal}, the risk is greatest during the first $1-2$ days. The prevalence and importance of aetiological factors for neonatal seizures are continuously changing and differ between developed and developing countries depending on the available care in NICUs. Among the numerous aetiological factors, hypoxic-ischaemic encephalopathy is the most common, especially among term neonates \cite{Panayiotopoulos_1970}. Epidemiology of neonatal seizures shows a high incidence rate in low-income settings \cite{mwaniki2010neonatal,pisani2018incidence}. For example, Sri Lankan data reveals that $3$ per $1000$ live births in term neonates and $7.5$ per $1000$ births in preterm neonates experience seizures. Since there are significantly fewer or no facilities for EEG monitoring in almost all of the Sri Lankan NICUs, these figures are likely to be under-reporting\cite{wanigasinghe_2017}.

Detecting neonatal seizures is particularly challenging because they often manifest subtly and can be mistaken for normal physiological behaviors. Therefore, having an objective monitoring method is critical. The gold standard method is video-EEG monitoring, which requires continuously monitoring the infant's brain activity using an EEG during suspected seizure events. This method, while reliable, is resource-intensive and may not always be feasible in resource limited clinical settings, mainly due to the unavailability of suitable equipment, the lack of experienced neonatologists and neurophysiologists for patient monitoring; ultimately contributing to critical causes for delays in the diagnosis process \cite{young2009continuous}. Therefore, several studies have been done in the past years to replace this monitoring task with machine learning and deep learning. Despite their reported performance, none of these have been integrated into hospital settings.

Among the earliest works, Temko \textit{et al.}~\cite{TEMKO2011464} designed a support vector machine classifier for a dataset from Cork University Maternity Hospital. With the recent advancements in deep learning, deep convolutional neural networks (CNN), recurrent neural networks, and long short-term memory, several studies have been carried out to classify EEG signals. In \cite{CNN_for_seizure,hossain}, authors have applied $2$D convolutions to detect seizures where in~\cite{hossain}, the input EEG signal is treated as a $2$D image and in~\cite{CNN_for_seizure}, the input is the spectrogram of the EEG epoch. Recently, models such as STATENET~\cite{Statenet} and ST-GAT~\cite{ST-GAT} have been introduced, where the temporal and spatial features are considered for the model prediction. The main drawback of these existing models is not being scalable to a reduced number of channels, which becomes an important requirement for neonatal seizure detection using low-cost hardware accessible to resource limited environments. Further, the model convergence is very slow and not able to explain the output concerning the particular EEG channels and time intervals of the input EEG epoch. Michele \textit{et al.}~\cite{GRAD_blink} introduced an explainable deep learning model for blink detection from EEG using gradient-weighted class activation mapping (Grad-cam)~\cite{GRADCAM} method which is capable of showing exactly where the blink occurs in the EEG epoch. In addition, several other studies in self-supervised learning (SSL)~\cite{ssl_1,SSL_2} were conducted to detect seizures. However, SSL-based methods provide slightly poor performance compared to the state-of-the-art (SOTA), such as an $8\%$ reduction in AUC. 

In this work, we introduce a new explainable deep learning model architecture that is capable of detecting seizures from EEG signals from a reduced EEG montage and interpreting the results in real time. As the CNN is a dominant architecture in computer vision tasks \cite{Alexnet} and sequence transduction tasks \cite{cnnSeqSeq}, our work leverages a CNN encoder, where $1$D convolutions are performed to extract temporal features, a graph attention (GAT) network for spatial feature extraction, and a binary classification head to classify EEG signals to seizure and normal states. For $80\%$ training data and $20\%$ test data, the model achieves an absolute improvement of $2.71\%$ and $16.33\%$ in AUC and recall, respectively. Moreover, when evaluated with a $10$-fold cross-validation, the model achieves an absolute improvement of $8.31\%$ and $42.86\%$ in AUC and recall respectively.
\section{Proposed Model Architecture}
This section is divided into two sections. In section~\ref{sec: Algo}, the proposed deep learning architecture is introduced in detail. To this end, we discuss the CNN encoder followed by the GAT network. In section~\ref{sec: explain}, model interpretability is discussed.
\subsection{Deep Learning Model Architecture}\label{sec: Algo}

\begin{figure*}[!ht]
\centering
\hspace{70pt}
  \begin{subfigure}{0.2\textwidth}
    \centering
    \includegraphics[width=\linewidth]{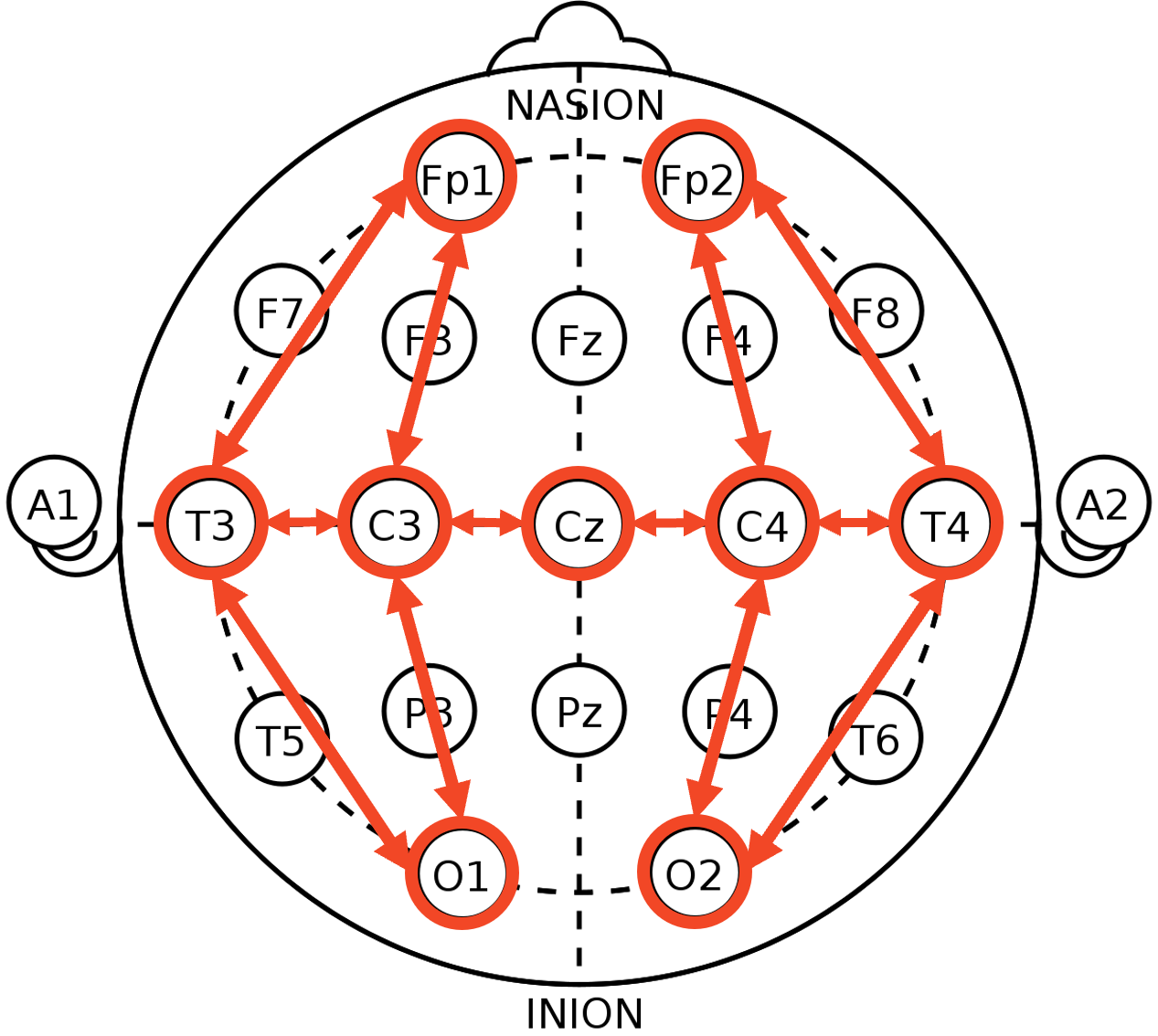}
    \caption{}
    \label{fig:electrode montage}
  \end{subfigure}
  \hfill
  \begin{subfigure}{0.39\textwidth}
    \centering
    \includegraphics[width=\linewidth]{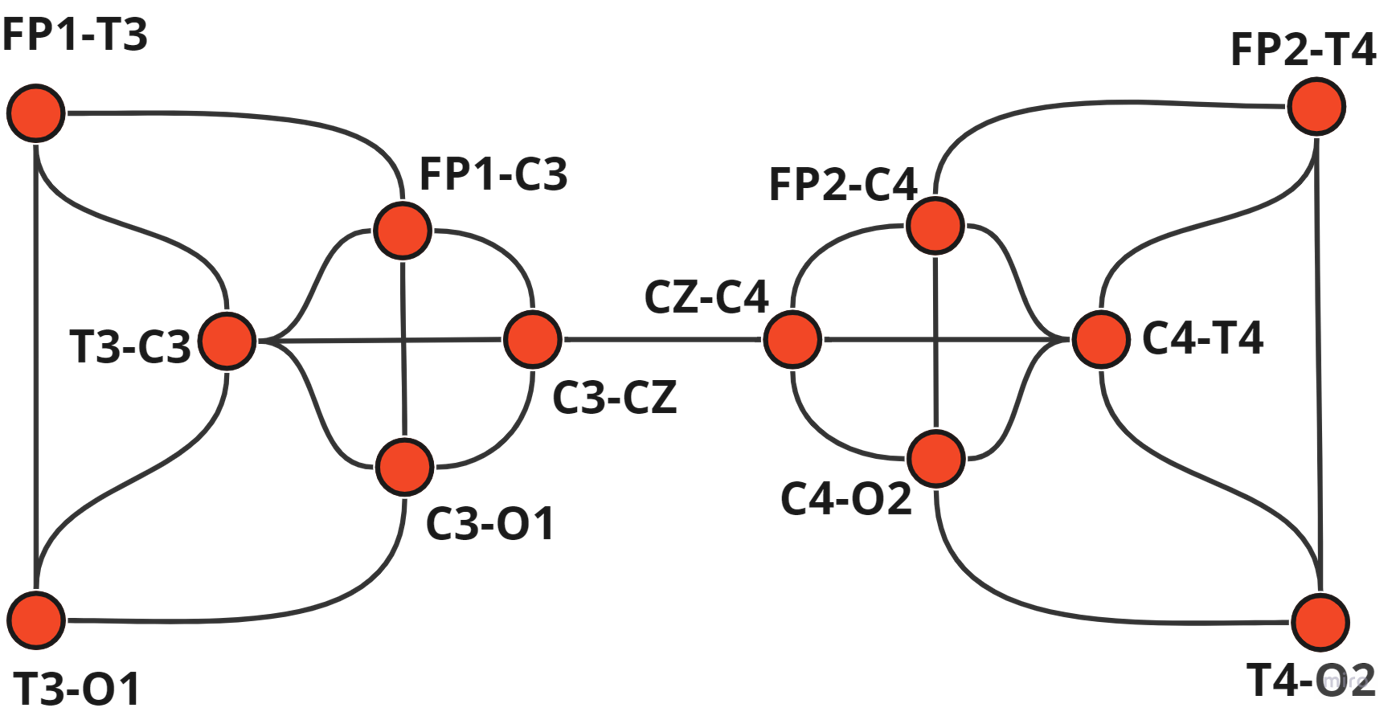}
    \caption{}
    \label{fig:graph}
  \end{subfigure}
  \hspace{20pt}
  \caption{(a) The proposed reduced montage electrode placement for seizure detection on the international $10-20$ system (b) The illustration of the employed reduced montage graph representation of the selected electrode montage. The graph nodes represent the channels and the edges represent the functional connectivity between channels.}
  \vspace{-10pt}
\end{figure*}

In this section, we introduce the proposed novel deep-learning model for real-time seizure detection from neonatal EEG signals. The proposed model employs $1)$ a CNN encoder~\ref{sec:CNN}, $2)$ a GAT network~\ref{sec:GAT}, and $3)$ a fully connected classification head~\ref{sec:classification}. The CNN encoder is used to extract the temporal features from the EEG epochs and the graph attention encoder is used to extract spatial features from the output of the CNN encoder. Apart from seizure detection, notably, we integrate interpretability in our model by leveraging a modified Grad-cam \cite{GRADCAM}, to explain which time ranges in each channel of a given EEG epoch contribute more significantly to the respective binary class of the model output. 

\subsubsection{CNN Encoder}\label{sec:CNN}
As the EEG signals are time series data, we use 1-D convolutions to extract the temporal features from EEG epochs. The CNN encoder employs four blocks, where each block utilizes convolutional layers with $(1\times5)$ and $(1\times7)$ receptive fields and \{$32,64,8,1$\} filters as shown in Fig.~\ref{fig:CNN}. After pre-processing the raw EEG data, the input matrix into this CNN encoder has the shape $12\times384$, where $12$ denotes the number of EEG channels and $384$ denotes the number of data samples within a time window of $12$ s.

Consider $F(x,\{W_{1i}\})$ and $H(x,\{W_{2i}\})$ as two different mapping functions of a set of stacked layers that output two different matrices with the same dimension given the same input matrix $x$. Therefore, we are able to do two mappings parallelly and add them together to obtain a new matrix with completely different features as in the equation:
\begin{equation}
H(x,\{W_i\})=F(x,\{W_{1i}\})+H(x,\{W_{2i}\})
\label{eq:1}
\end{equation}
This technique is applied in the convolutional Block $1$ as seen in Fig.~\ref{fig:CNN}, to extract different temporal features simultaneously by applying different kernel sizes in two parallel convolutional layers. Block $1$ is followed by another $3$ convolutional blocks, each of them having a residual learning framework as proposed in \cite{resnet} for fast convergence of the model. In this CNN encoder, the skip connections simply perform identical mapping to preserve the original input dimensions when adding. 

Since this CNN encoder is a tiny network with only $8$ convolution layers, it proves difficult to achieve a good training performance with a simple sequential network. In order to introduce non-linearity in our model, we incorporated the widely used rectified linear unit ($ReLU$) as the activation of each convolution layer, since other activation functions may vanish the gradients in backpropagation or result in higher training time. We further experimented with the $Swish$ activation function, however, that increased the training time by approximately $13$ minutes compared to the $ReLU$ function, which did not seem beneficial.

After adding convolution outputs in each block, average pooling is performed to downsample the feature map by a factor of $2$ for better optimization of the model and batch normalization as a solution to the gradient exploding issue according to \cite{batch}. The reason for using average pooling instead of max pooling is to aggregate more temporal information into one feature point rather than solely depending on a single value within a moving window. This CNN encoder is designed such that it reduces overfitting and training time, and stops gradient degradation with the help of residual and parallel connections. The selection of the receptive fields, number of filters, and layers is decided by a rigorous ablation study.

\subsubsection{GAT Network}\label{sec:GAT}
\textbf{Graph Representation}: After extracting the temporal features from a signal, we need to extract spatial features from the EEG epochs with the help of interchannel connectivity. Here, we employ a graph GAT network \cite{ST-GAT, GAT} to extract the spatial feature. The requirements to generate the graph are the vertices that correspond to the 12 channels, their feature vectors which are the output features from the CNN encoder for each channel, and the adjacency matrix to denote the functional connectivity between channel pairs. Further, the selected channels given in section~\ref{sec: dataset}, have the capability of modeling brain connectivity as in the graph in Fig.~\ref{fig:graph}. 

According to Tekgul \textit{et al.} in \cite{Electrode}, the localization of EEG seizures with reduced electrode montage is acceptable and compared to a standard $10-20$ EEG system as most neonatal seizures occur in the central zone of the brain. Hence, not considering global inter-hemisphere connections in the Front Lobe and Parietal Lobe would not reduce the network efficiency. Additionally, other neonatal seizures occur in bilateral posterior and anterior regions which are covered by O1, O2, and Fp1, Fp2, respectively. Also, due to the EEG channels T3-C3, C3-CZ, CZ-C4, and C4-T4, the biological connection between the left and right hemispheres is maintained throughout the process. Therefore, with the designed graph, it is possible to leverage the information passing between left and right parieto-occipital or fronto-temporal zones. As neonates have very small brains, the proposed graph rarely misses the event of a seizure \cite{stevenson2018effect}. 

\begin{figure}[!t]
    \vspace{-10pt}
    \centering
    \includegraphics[scale=0.3]{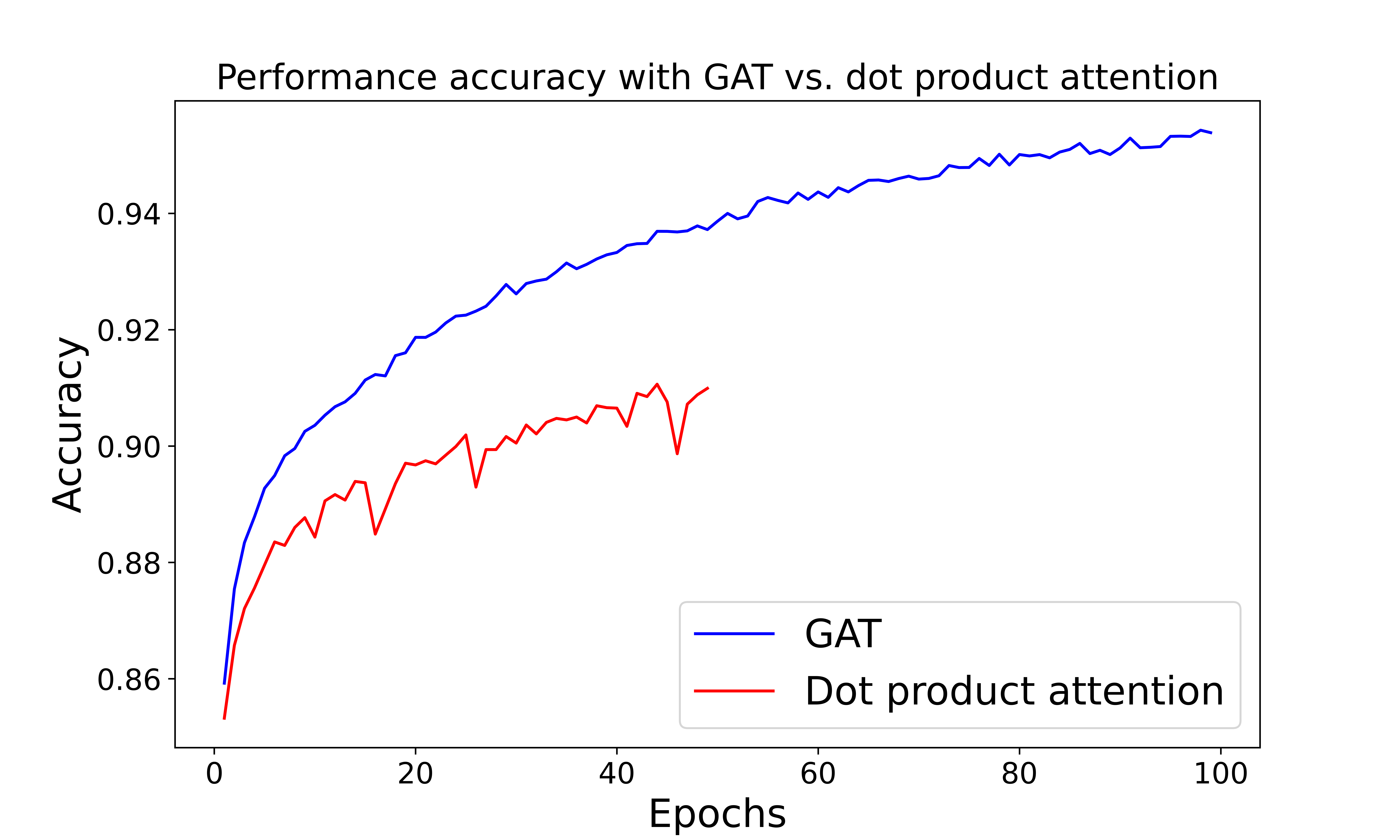}
    \caption{Performance comparison between GAT layers and scaled dot product attention layers. Training for dot product attention was terminated after $50$ epochs due to low performance.}
    \label{fig:GATvsDOT}
    \vspace{-20pt}
\end{figure}

\textbf{Attention Layers}: It is imperative to pay attention to the connected EEG channels in the selected electrode montage when extracting spatial features. From two widely used approaches for building an attention mechanism, $1$) a network built with GAT layers \cite{GAT} or $2$) a network built with scaled dot-product attention \cite{transformer}, we opted for a GAT as it slightly outperforms scaled dot-product attention as shown in Fig.~\ref{fig:GATvsDOT}. In addition to this, the fact that the brain network could be modeled as a graph motivated us to apply a GAT network to extract spatial features.

In a GAT layer, each node aggregates features from adjacent nodes and constructs a new feature set for itself. Given the feature sets for each node; $\textbf{\textit{H}} {\in}{\mathbb{R}^{12\times F}}$, a learnable weight matrix; $\textbf{\textit{W}} {\in}{\mathbb{R}^{F\times F^{'}}}$ is required to linearly transform the input features to high-level features by the simple matrix multiplication; $\textbf{\textit{H}}\times \textbf{\textit{W}} {\in}{\mathbb{R}^{12\times F^{'}}}$. Next, a shared masked self-attention is performed to compute the attention coefficients and masking is done according to the adjacency matrix. The eq.~\ref{eq:2} from Peter \textit{et al.} \cite{GAT}, explains how to compute the attention coefficients from the $j^{th}$ node to the $i^{th}$ node ($\alpha_{ij}$)
\begin{equation}
\alpha_{ij}=\frac{\text{exp(LeakyReLU(}[\textbf{\textit{h}}_{i}\textbf{\textit{W}}||\textbf{\textit{h}}_{j}\textbf{\textit{W}}]\textbf{\textit{a}}^T))}{\sum_{k{\in}\mathcal{N}_{i}}\text{exp(LeakyReLU(}[\textbf{\textit{h}}_{i}\textbf{\textit{W}}||\textbf{\textit{h}}_{k}\textbf{\textit{W}}]\textbf{\textit{a}}^T))}
\label{eq:2}
\end{equation}

Here, $\textbf{\textit{h}}_{i}$ denotes a row of $\textbf{\textit{H}}$ matrix, $\mathcal{N}_i$ denotes all the neighbor nodes of node $i$ and itself, $\textbf{\textit{a}}{\in}\mathbb{R}^{2F^{'}}$ is a learnable weight vector, and $||$ represents concatenation. Once these attention coefficients are obtained, the new features set for each node are calculated by a simple non-linear transformation; 
\begin{equation}
    \textbf{\textit{H}}^{'}=ELU(\textbf{\textit{AHW}})
\end{equation} 
where $\textbf{\textit{A}}{\in}{\mathbb{R}}^{12\times 12}$ is the masked attention coefficient matrix. 

This mask makes sure that a node pays attention only to itself and its first-order neighbors. Hence, we apply $3$ GAT layers after the CNN encoder, to achieve an optimal spatial receptive field by aggregating features from the $3^{rd}$-order neighbors which will efficiently cover $78\%$ of the brain network. If one or two more layers are applied, the receptive field will be increased, however, it will not improve the network efficiency and model performance. Therefore it was proved experimentally that the optimal number of GAT layers should be $3$. Here, as mentioned in Fig.~\ref{fig:CNN}, the output feature maps of the GAT layers $1, 2,$ and $3$ have the shapes $(12\times 37)$, $(12\times 32)$, and $(12\times 16)$, respectively. In their work, Raeis \textit{et al.} \cite{ST-GAT} adopted a similar approach for feature map selection, while employing 18 EEG channels.

\subsubsection{Classification Head}\label{sec:classification}
The classification task is performed through a multilayer perception (MLP) after extracting temporal and spatial features from the proposed CNN encoder and GAT network. This network consists of $3$ dense layers of $32, 16,$ and $1$ neurons, respectively. The first and second dense layers are followed by the $ReLU$ function while the final layer is followed by the $Sigmoid$ function. A global average pooling layer is applied along the temporal axis before these in order to reduce the GAT output dimensions to $(12\times1)$.

\subsection{Model Interpretability}\label{sec: explain}
High transparency is essential in deep learning applications in medicine. This emphasizes the relevance of explainability in medical AI. To this end, this work proposes a new approach leveraging Grad-cam \cite{GRADCAM}, in which the gradient of the class activation (logit value) is obtained with respect to the activations of the last GAT layer to generate a heatmap of the shape of the input signal. Once the gradient computation is complete, the final GAT layer outputs are scaled by the mean of gradients. The resulting values are passed through a $ReLU$ activation function and then normalized using min-max normalization. This generates a heatmap, where $0$ represents the least relevance and $1$ represents the highest relevance to the output. This heatmap is then mapped to a standard colormap to visualize the dependence of the relevance of specific time periods in each of the $12$ channels of a given EEG epoch to the respective binary class of the model output. This is clearly visualized in Fig.~\ref{fig:exAI} with a blue-white-red (bwr) colormap. Further, this step does not affect the deep learning model binary class output as it runs as a post-classification task.

\section{Dataset and Pre-processing}\label{sec: dataset}
In this study, the publicly available Helsinki Zenodo scalp EEG dataset \cite{stevenson2019dataset} is used to train and test the deep learning model. This open-source dataset contains $74$-min (median) long multi-channel EEG recordings, sampled at $256$ Hz, from $79$ term neonates admitted to the NICU at Helsinki University Hospital, Finland. It consists of $3$ annotation files created by $3$ independent trained neurologists. As a result, only $39$ neonates were identified as having seizures by consensus, while $22$ were identified as seizure-free.

This dataset, recorded with respect to a reference point, allows for the construction of a number of EEG channels. According to the American Clinical Neurophysiology Society recommendations, the electrode placement should follow the international $10$-$20$ system modified for neonates. Consequently, most existing methods, including the state-of-the-art, utilize $18$ EEG channels for training and evaluation. Even though the full array is recommended, the Minimum Technical Standards for Pediatric Electroencephalography states that it is acceptable to use a reduced array wherever necessary \cite{shellhaas2011american,kuratani2016american}. Hence, our study employs just $12$ channels, selected based on Tekgul \textit{et al.} \cite{Electrode} to model the double banana-shaped reduced electrode montage. The specific channels used are, Fp$1$-T$3$, T$3$-O$1$, Fp$1$-C$3$, C$3$-O$1$, Fp$2$-C$4$, C$4$-O$2$, Fp$2$-T$4$, T$4$-O$2$, T$3$-C$3$, C$3$-CZ, CZ-C$4$, C$4$-T$4$ as shown in Fig.~\ref{fig:electrode montage}.

In EEG signal processing, it is important to preprocess the signals to remove noise and artifacts. For this dataset, flat lines at $0$ V in some EEG signals need to be removed before further processing. An automated procedure is implemented to identify and remove these flat lines from the signals. Additionally, a bandpass Chebyshev type-II digital filter with $1$ Hz and $16$ Hz cutoff frequencies is applied to eliminate baseline drift and high-frequency noise components. The original sampling frequency of $256$ Hz is down-sampled to $32$ Hz to reduce model complexity and training times. Finally, the EEG epochs are normalized to scale the signal amplitudes.

We select only the $39$ neonates with seizures identified by consensus to train and evaluate our model as the other $22$ neonatal seizure-free signals cause huge class imbalance issues and the remaining $18$ neonatal signals have the potential to be misclassified as actual seizures. Although neonates with seizures by consensus are selected, there is a significant difference in total seizure duration compared to non-seizure duration. The total seizure duration represents only $18.14\%$ of the total EEG signal duration. To address this class imbalance issue, we adopted a technique from \cite{ST-GAT}, where we overlap each 12-second epoch with $11$ seconds for seizure segments and 10 seconds for non-seizure segments. While this approach does not entirely eliminate the class imbalance, it effectively reduces its impact on displaying a seizure-to-non-seizure epochs ratio of $1:2$. The remaining class imbalance issue is mitigated by applying focal binary cross-entropy loss in training.

\section{Model Training}
\label{sec:Training and hyperparameter tuning}
\begin{table*}[!ht]
\scriptsize
  \centering
  \caption{Model performance comparison. CV - Cross Validation}
  \label{tab:Result_table}
  \begin{tabular}{|c|c|c|c|c|c|c|c|}
    \hline
    Number of EEG & Method & Accuracy $(\%)$& \multicolumn{2}{c|}{AUC$(\%)$} & Recall & Precision & Kappa\\ \cline{3-5}
     channels& & mean$\pm$ std & Median (IQR)&Mean$\pm$std&$(\%)$ &$(\%)$ &\\ \hline
    &MSC-GCNN\cite{12}&-&$99.10$ $(96.80,99.60)$ &$94.70\pm10.90$&$96.71$&-&$0.80$\\ \cline{2-8}
    $18^{1}$&PLV-GCNN\cite{12}&-&$99.00$ $(95.20,99.70)$ &$94.10\pm10.50$&$95.30$&-&$0.79$\\ \cline{2-8}
    &SD-GCNN\cite{12}&-&$97.30$ $(86.30,99.60)$ &$90.09\pm13.50$&$96.68$&-&$0.71$\\ \cline{2-8}
    &ST-GAT(FL)\cite{ST-GAT}&-&$\mathbf{99.30}$ $(96.40,99.50)$ &$\mathbf{96.60}$$\pm$$8.90$&$\mathbf{98.00}$&-&$0.88$\\ \hline
   $12$ &ST-GAT (FL)$^2$&$80.29$$\pm$$9.48$&$83.98$ $(77.80,90.90)$ &$83.15\pm8.85$&$39.98$&$\mathbf{94.91}$&$0.43$\\ \cline{2-8}
    $10$-fold CV&Our method&$\mathbf{89.02}\pm2.91$&$\mathbf{91.84}$ $ (88.57,95.21)$ &$\mathbf{91.46}\pm4.36$&$\mathbf{82.84}$&$94.23$&$\mathbf{0.89}$\\ \hline
    $12$ &ST-GAT (FL)$^2$&$88.80$&\multicolumn{2}{c|}{$91.71$}&$66.89$&$\mathbf{95.17}$&$0.71$\\ \cline{2-8}
    $(80\%-20\%)$&Our method&$\mathbf{91.56}$&\multicolumn{2}{c|}{$\mathbf{94.42}$ }&$\mathbf{83.22}$&$88.61$&$\mathbf{0.80}$\\ \hline
    \multicolumn{8}{l}{}\\
    \multicolumn{8}{l}{$^{1}$These models were trained with $18$-channel full montage EEG data. Since no prior study had been conducted using the $12$-channel reduced}\\
    \multicolumn{8}{l}{montage, we report these here to demonstrate that our model performs better than SOTA methods in terms of Cohen's kappa while having a}\\
    \multicolumn{8}{l}{reduced number of channels.} \\
    \multicolumn{8}{l}{$^2$We retrained ST-GAT (FL), the best variant of the ST-GAT, for the $12$-channel reduced montage and reported the evaluated results to show a}\\
    \multicolumn{8}{l}{fair comparison between our model and the current best SOTA model.}\\
  \end{tabular}
    \vspace{-15pt}
\end{table*}

The model is trained and evaluated in two approaches; $1$) allocating randomly selected 31 subjects for the training data set ($\sim80\%$) and the remaining $8$ subjects for the test dataset ($\sim20\%$) from $39$ neonates with seizures by consensus and $2$) performing $10$-fold cross-validation on these $39$ neonates. The model is trained with the Adam optimizer with a $0.002$ learning rate and focal binary cross entropy loss with $\gamma=2$ and $\alpha=0.4$ as the loss function to address the class imbalance issue between seizure and non-seizure samples. We apply a dropout of $0.2$ probability at the end of each convolutional Block $1, 2, 3,$ and between every GAT layer and dense layer. Additionally, a $L_2$ kernel regularizer with a $0.0001$ regularization value is selected. These hyperparameters were selected after an ablation study. 

\section{Results and Discussion}
\label{sec:Results}

Table~\ref{tab:Result_table} compares the performance results of previously published models with our proposed method. These comparisons are inequitable due to the fact that previously published methods incorporate 18 EEG channels, which are obtained from the full electrode montage. In order to ensure a fair comparison with the SOTA, we retrained the best performing model among the current SOTA, ST-GAT (FL) which is the best variant of ST-GAT as in~\cite{ST-GAT}, for the 12-channel reduced montage and reported the results. Our model performs better when evaluated with a $10$-fold cross-validation, where the model achieves an absolute improvement of $8.31\%$ and $42.86\%$ in mean AUC and recall, respectively. Further, for $80\%$ training data and $20\%$ test data, the model achieves an absolute improvement of $2.71\%$ and $16.33\%$ in mean AUC and recall, respectively. Further, our model evaluated with $10$-fold cross-validation, has the highest Cohen's kappa value regardless of the reduced montage.

Our model not only detects seizures in EEG epochs but also provides interpretable outputs for real-time analysis. This highlights the specific channels and time windows that are most critical for the model's decision. By visualizing these regions using a bwr colormap as in Fig.~\ref{fig:exAI}, we can assess the model's ability to understand and interpret the input. The first subplot of Fig.~\ref{fig:exAI} shows the true labels and predicted seizure probability for a $7.5$-minute long EEG signal. As shown in the next $12$ subplots, our model effectively differentiates between seizure activity (from $4$ mins $25$ secs onwards) and the seizure-free period (up to $4$ mins $25$ secs). This demonstrates the model's capability to distinguish between these two crucial elements in EEG seizure analysis even though the artifacts in the seizure-free region are in the same amplitude range as seizures. Since the understanding of how the model identifies a given EEG epoch as a seizure epoch is more important, we visualize the heatmaps of only EEG epochs detected as seizure by the model otherwise, we visualize EEG epochs in blue color. The last subplot of Fig.~\ref{fig:exAI}, a zoomed-in time window of the Fp$1$-T$3$ channel, shows the onset of the seizure which demonstrates how specific channels and time windows of the EEG recording were critical to detect the occurrence of the seizure. Further, real-time results can be found in the git-hub repository: \url{https://github.com/Dinuka-1999/BraiNeoCare}
\begin{figure}[!t]
    \centering
    \vspace{-0.1in}
    \includegraphics[width=\linewidth,height=5in]{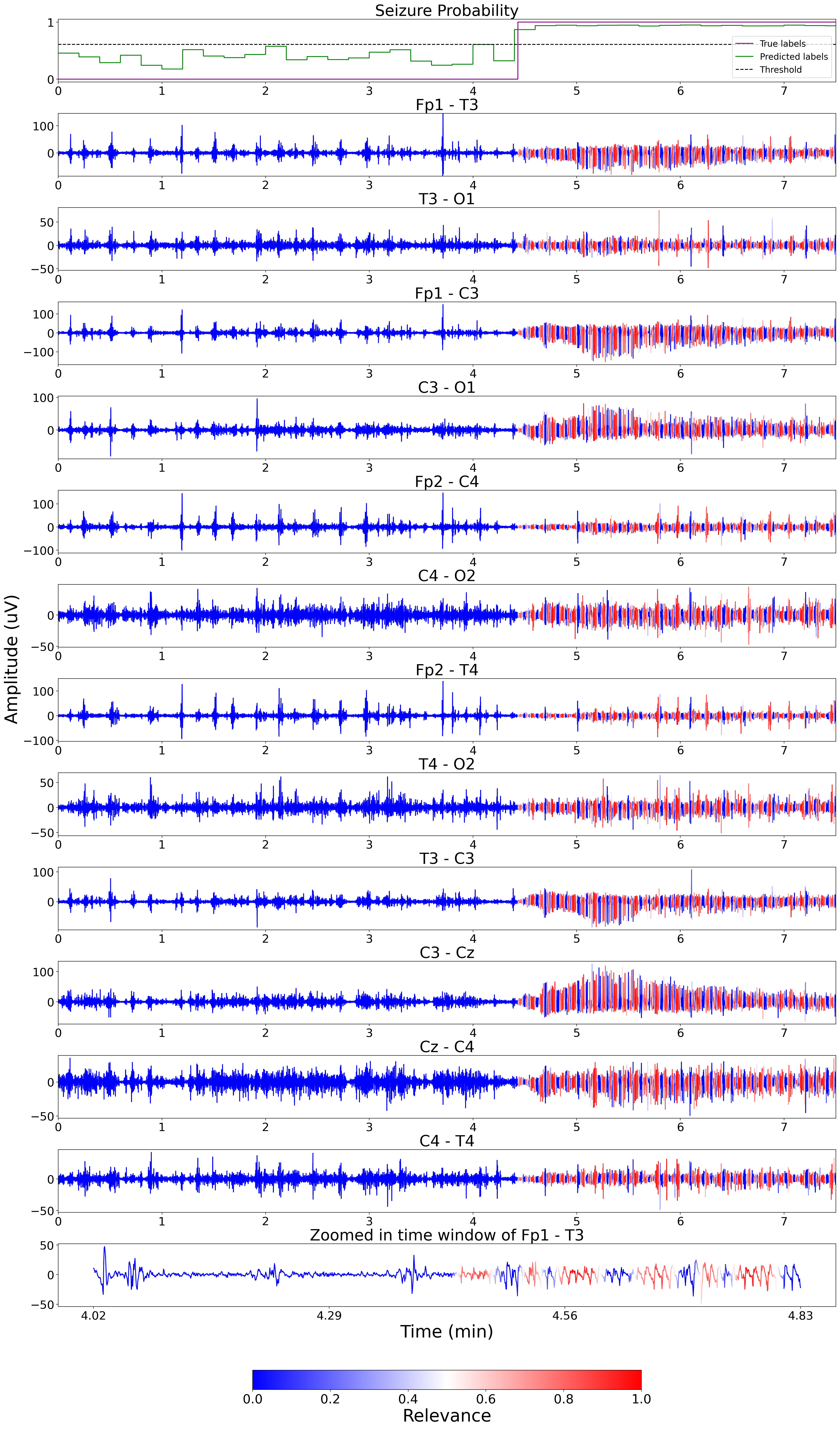}
    \caption{\textbf{Top subplot:} Comparison between true labels and model prediction probabilities.  \textbf{Next $12$ subplots:} Visualization of $7.5$ minutes EEG, where the recording is observed to be seizure free up to $4$ mins $25$ secs, and a seizure occurs past this point. \textbf{Last subplot:} A zoomed-in version for better visualization of seizure onset and how the relevance changes.}
    \label{fig:exAI}
    \vspace{-15pt}
\end{figure}


\section{Conclusion }
\label{sec:Conclusion}

Although neonatal seizure detection is a challenging task even for experienced professionals, by introducing well-defined signal processing techniques and deep learning models, we may attempt to make the task less challenging. By leveraging explainable AI and a seizure detection probability distribution, not only experienced professionals, but less experienced professionals can gain the skillset to diagnose seizure events accurately and provide prompt management. To this end, we have presented an efficient, reliable, and unique deep learning architecture built upon a CNN encoder to extract temporal features, a GAT network to extract spatial features, and a binary classification head. On average, it takes only $62$ ms to detect seizures in a $12$-second EEG epoch on the CPU, with even faster processing ($32$ ms) on the GPU. Beyond its ability to detect seizures in real time with a reduced montage, this model offers the unique advantage of real-time interpretability. This allows for quick and insightful analysis. A modified version of Grad-cam is employed to explain the model's binary class output demonstrating which channels and time windows have been looked at by the model when a seizure is detected. The reduced montage employs only $9$ electrodes making it easy to prepare the subject for testing and increasing the patient comfortability.

For future work, we see great potential for the medical field in improving the model's performance with real-time fast artifact removal, and embedded machine learning for real-time seizure detection. An important task is to improve model performance by self-supervised training with a large, unlabeled EEG dataset. To accomplish this, as our next step, we are planning to test this proposed trained model with data collected under the supervision of trained experts from The Lady Ridgeway Hospital in Colombo, Sri Lanka.

\section*{Acknowledgement}

The authors would like to extend gratitude to Dr. Ranga Rodrigo for providing computational resources, which were funded by the Accelerating Higher Education Expansion and Development (AHEAD) grant. The authors would also like to thank the staff of The Lady Ridgeway Hospital in Colombo and the Department of Paediatrics, Faculty of Medicine, University of Colombo for their dedication and support to Neonatal Care in Sri Lanka.






\begin{thebibliography}{99}
\bibitem{kang2015neonatal}S. K. Kang, and S. D. Kadam, ``Neonatal seizures: impact on neurodevelopmental outcomes," \textit{frontiers in Pediatrics}, vol. 3, p. 101, 2015.
\bibitem{Panayiotopoulos_1970}C. Panayiotopoulos, ``Neonatal seizures and neonatal syndromes," in \textit{The epilepsies: seizures, syndromes and management}, Bladon Medical Publishing, 2005.
\bibitem{mwaniki2010neonatal}M. Mwaniki, A. Mathenge, S. Gwer, N. Mturi, E. Bauni, C. R. Newton, J. Berkley, and R. Idro, ``Neonatal seizures in a rural Kenyan District Hospital: aetiology, incidence and outcome of hospitalization." \textit{BMC medicine}, vol. 8, pp. 1-8, 2010.
\bibitem{pisani2018incidence}F. Pisani, C. Facini, E. Bianchi, G. Giussani, B. Piccolo, and E. Beghi, ``Incidence of neonatal seizures, perinatal risk factors for epilepsy and mortality after neonatal seizures in the province of Parma, Italy," \textit{Epilepsia}, vol. 59, no. 9, pp. 1764-1773, 2018.
\bibitem{wanigasinghe_2017}J. Wanigasinghe, R. Kapurubandara, C. Arambepola, S. Sri Ranganathan, and J. Philips, ``Incidence of neonatal seizures in babies born in two premier maternity hospitals within Colombo city," in \textit{Association of Sri Lankan Neurologists 11th Annual Academic Sessions}, pp. 43-44, 2017.
\bibitem{young2009continuous}G. B. Young, ``Continuous EEG monitoring in the ICU: challenges and opportunities." \textit{The Canadian Journal of Neurological sciences. Le Journal Canadien des Sciences Neurologiques}, vol. 36, pp. S89-91, 2009.
\bibitem{TEMKO2011464}A. Temko, E. Thomas, W. Marnane, G. Lightbody, and G. Boylan, ``Eeg-based neonatal seizure detection with support vector machines,” \textit{Clinical Neurophysiology}, vol. 122, no. 3, pp. 464–473, 2011.
\bibitem{hossain}M. S. Hossain, S. U. Amin, M. Alsulaiman, and G. Muhammad, ``Applying deep learning for epilepsy seizure detection and brain mapping visualization,” \textit{ACM Transactions on Multimedia Computing, Communications, and Applications (TOMM)}, vol. 15, no. 1s, pp. 1–17, 2019.
\bibitem{CNN_for_seizure}P. Thodoroff, J. Pineau, and A. Lim, ``Learning robust features using deep learning for automatic seizure detection,” in \textit{Machine learning for healthcare conference}. PMLR, 2016, pp. 178–190.
\bibitem{Statenet}Z. Li, Y. Fang, Y. Li, K. Ren, Y. Wang, X. Luo, J. Duan, C. Huang, D. Li, and L. Qiu, ``Protecting the future: Neonatal seizure detection with spatial-temporal modeling,” in \textit{2023 IEEE International Conference on Systems, Man, and Cybernetics (SMC)}. IEEE, 2023, pp. 196–201
\bibitem{ST-GAT}K. Raeisi, M. Khazaei, G. Tamburro, P. Croce, S. Comani, F. Zappasodi \textit{et al}., ``A class-imbalance aware and explainable spatio-temporal graph attention network for neonatal seizure detection,” \textit{International Journal of Neural Systems}, vol. 33, no. 9, p. 2350046, 2023.
\bibitem{GRAD_blink}M. L. Giudice, N. Mammone, C. Ieracitano, M. Campolo, A. R. Bruna, V. Tomaselli, and F. C. Morabito, ``Visual explanations of deep convolutional neural network for eye blinks detection in eeg-based bci applications,” in \textit{2022 International Joint Conference on Neural Networks (IJCNN)}. IEEE, 2022, pp. 01–08.
\bibitem{GRADCAM}R. R. Selvaraju, M. Cogswell, A. Das, R. Vedantam, D. Parikh, and D. Batra, ``Grad-cam: Visual explanations from deep networks via gradient-based localization,” in \textit{Proceedings of the IEEE international conference on computer vision}, 2017, pp. 618–626
\bibitem{ssl_1}S. Das, P. Pandey, and K. P. Miyapuram, ``Improving self-supervised pretraining models for epileptic seizure detection from eeg data,” \textit{arXiv preprint arXiv:2207.06911}, 2022.
\bibitem{SSL_2}D. Cai, J. Chen, Y. Yang, T. Liu, and Y. Li, ``Mbrain: A multi-channel self-supervised learning framework for brain signals,” in \textit{Proceedings of the 29th ACM SIGKDD Conference on Knowledge Discovery and Data Mining}, 2023, pp. 130–141.
\bibitem{Alexnet}A. Krizhevsky, I. Sutskever, and G. E. Hinton, ``Imagenet classification with deep convolutional neural networks,” \textit{Advances in neural information processing systems}, vol. 25, 2012.
\bibitem{cnnSeqSeq}J. Gehring, M. Auli, D. Grangier, D. Yarats, and Y. N. Dauphin, ``Convolutional sequence to sequence learning,” in \textit{International conference on machine learning}. PMLR, 2017, pp. 1243–1252.
\bibitem{resnet}K. He, X. Zhang, S. Ren, and J. Sun, ``Deep residual learning for image recognition,” in \textit{Proceedings of the IEEE conference on computer vision and pattern recognition}, 2016, pp. 770–778.
\bibitem{batch}S. Ioffe and C. Szegedy, ``Batch normalization: Accelerating deep network training by reducing internal covariate shift,” in \textit{International conference on machine learning}. pmlr, 2015, pp. 448–456.
\bibitem{GAT}P. Veličković, G. Cucurull, A. Casanova, A. Romero, P. Liò, and Y. Bengio, ``Graph attention networks,” \textit{in International Conference on Learning Representations}, 2018.
\bibitem{Electrode}H. Tekgul, B. F. Bourgeois, K. Gauvreau, and A. M. Bergin, ``Electroencephalography in neonatal seizures: comparison of a reduced and a full 10/20 montage,” \textit{Pediatric neurology}, vol. 32, no. 3, pp. 155–161, 2005.
\bibitem{stevenson2018effect}N. J. Stevenson, L. Lauronen, and S. Vanhatalo, ``The effect of reducing eeg electrode number on the visual interpretation of the human expert for neonatal seizure detection,” \textit{Clinical Neurophysiology}, vol. 129, no. 1, pp. 265–270, 2018.
\bibitem{transformer}A. Vaswani, N. Shazeer, N. Parmar, J. Uszkoreit, L. Jones, A. N. Gomez, Ł. Kaiser, and I. Polosukhin, ``Attention is all you need,” \textit{Advances in neural information processing systems}, vol. 30, 2017.
\bibitem{stevenson2019dataset}N. J. Stevenson, K. Tapani, L. Lauronen, and S. Vanhatalo, ``A dataset of neonatal eeg recordings with seizure annotations,” \textit{Scientific data}, vol. 6, no. 1, pp. 1–8, 2019.
\bibitem{shellhaas2011american}R. A. Shellhaas, T. Chang, T. Tsuchida, M. S. Scher, J. J. Riviello, N. S. Abend, S. Nguyen, C. J. Wusthoff, and R. R. Clancy, ``The american clinical neurophysiology society’s guideline on continuous electroencephalography monitoring in neonates,” \textit{Journal of clinical neurophysiology}, vol. 28, no. 6, pp. 611–617, 2011.
\bibitem{kuratani2016american}J. Kuratani, P. L. Pearl, L. R. Sullivan, R. M. S. Riel-Romero, J. Cheek, M. M. Stecker, D. S. J. Orta, O. Selioutski, S. R. Sinha, F. W. Drislane et al., ``American clinical neurophysiology society guideline 5: minimum technical standards for pediatric electroencephalography,” \textit{The Neurodiagnostic Journal}, vol. 56, no. 4, pp. 266–275, 2016.
\bibitem{12}K. Raeisi, M. Khazaei, P. Croce, G. Tamburro, S. Comani, and F. Zappasodi, ``A graph convolutional neural network for the automated detection of seizures in the neonatal eeg,” \textit{Computer methods and programs in biomedicine}, vol. 222, p. 106950, 2022.
\end{thebibliography}
\end{document}